%% file: main.tex
\documentclass[sigconf]{acmart}
\usepackage{float}

\input{preamble}
\begin{document}

\title{Advancing computational reproducibility in the \\Dataverse data repository platform} 


\input{authors}

\begin{abstract}
Recent reproducibility case studies have raised concerns showing that much of the deposited research has not been reproducible. One of their conclusions was that the way data repositories store research data and code cannot fully facilitate reproducibility due to the absence of a runtime environment needed for the code execution. New specialized reproducibility tools provide cloud-based computational environments for code encapsulation, thus enabling research portability and reproducibility. However, they do not often enable research discoverability, standardized data citation, or long-term archival like data repositories do. This paper addresses the shortcomings of data repositories and reproducibility tools and how they could be overcome to improve the current lack of computational reproducibility in published and archived research outputs.
\end{abstract}

\input{ccs}

\maketitle
\input{main-text}

\begin{acks}
Thank you to the Dataverse team for their help and support. This work is partially funded by the Sloan Foundation. Ana Trisovic is funded by the Sloan Foundation.
\end{acks}

\bibliographystyle{ACM-Reference-Format}
\bibliography{sample-base}

\end{document}

%% file: preamble.tex
\AtBeginDocument{%
  \providecommand\BibTeX{{%
    \normalfont B\kern-0.5em{\scshape i\kern-0.25em b}\kern-0.8em\TeX}}}

\setcopyright{acmcopyright}
\copyrightyear{2020}
\acmYear{2020}
\acmDOI{10.1145/1122445.1122456}

\acmConference[P-RECS'20]{3rd International Workshop on Practical Reproducible Evaluation of Systems (P-RECS'20)}{June 23, 2020}{Stockholm, Sweden}



%% file: authors.tex
\author{Ana Trisovic}
\affiliation{%
  \institution{Institute for Quantitative Social Science, Harvard University}
  \streetaddress{1737 Cambridge St}
  \city{Cambridge}
  \state{MA}
  \postcode{02138}
\country{USA}}
\email{anatrisovic@g.harvard.edu}

\author{Philip Durbin}
\affiliation{%
  \institution{Institute for Quantitative Social Science, Harvard University}
  \streetaddress{1737 Cambridge St}
  \city{Cambridge}
  \state{MA}
  \postcode{02138}
\country{USA}}

\author{Tania Schlatter}
\affiliation{%
  \institution{Institute for Quantitative Social Science, Harvard University}
  \streetaddress{1737 Cambridge St}
  \city{Cambridge}
  \state{MA}
  \postcode{02138}
\country{USA}}

\author{Gustavo Durand}
\affiliation{%
  \institution{Institute for Quantitative Social Science, Harvard University}
  \streetaddress{1737 Cambridge St}
  \city{Cambridge}
  \state{MA}
  \postcode{02138}
\country{USA}}

\author{Sonia Barbosa}
\affiliation{%
  \institution{Institute for Quantitative Social Science, Harvard University}
  \streetaddress{1737 Cambridge St}
  \city{Cambridge}
  \state{MA}
  \postcode{02138}
\country{USA}}

\author{Danny Brooke}
\affiliation{%
  \institution{Institute for Quantitative Social Science, Harvard University}
  \streetaddress{1737 Cambridge St}
  \city{Cambridge}
  \state{MA}
  \postcode{02138}
\country{USA}}

\author{Merc\`e Crosas}
\affiliation{%
  \institution{Institute for Quantitative Social Science, Harvard University}
  \streetaddress{1737 Cambridge St}
  \city{Cambridge}
  \state{MA}
  \postcode{02138}
\country{USA}}
\email{mcrosas@g.harvard.edu}

\renewcommand{\shortauthors}{A. Trisovic et al.}

%% file: ccs.tex
\ccsdesc[300]{Information systems~Digital libraries and archives}
\ccsdesc[300]{Information systems~Computing platforms}

%% file: main-text.tex
\section{Introduction}

The requirement of reproducible computational research is becoming increasingly important and mandatory across the sciences~\cite{national2019reproducibility}. Because reproducibility implies a certain level of openness and sharing of data and code, parts of the scientific community have developed standards around documenting and publishing these research outputs~\cite{wilkinson2016fair,chen2019open}. Publishing data and code as a replication package in a data repository is considered to be a best practice for enabling research reproducibility and transparency~\cite{molloy2011open}.~\footnote{At Dataverse, the data, and code used to reproduce a published study are called "replication data" or a "replication package"; in Whole Tale, this is called a "tale", and in Code Ocean a "reproducible capsule".}

\begin{figure}
  \centering
    \includegraphics[width=0.9\linewidth]{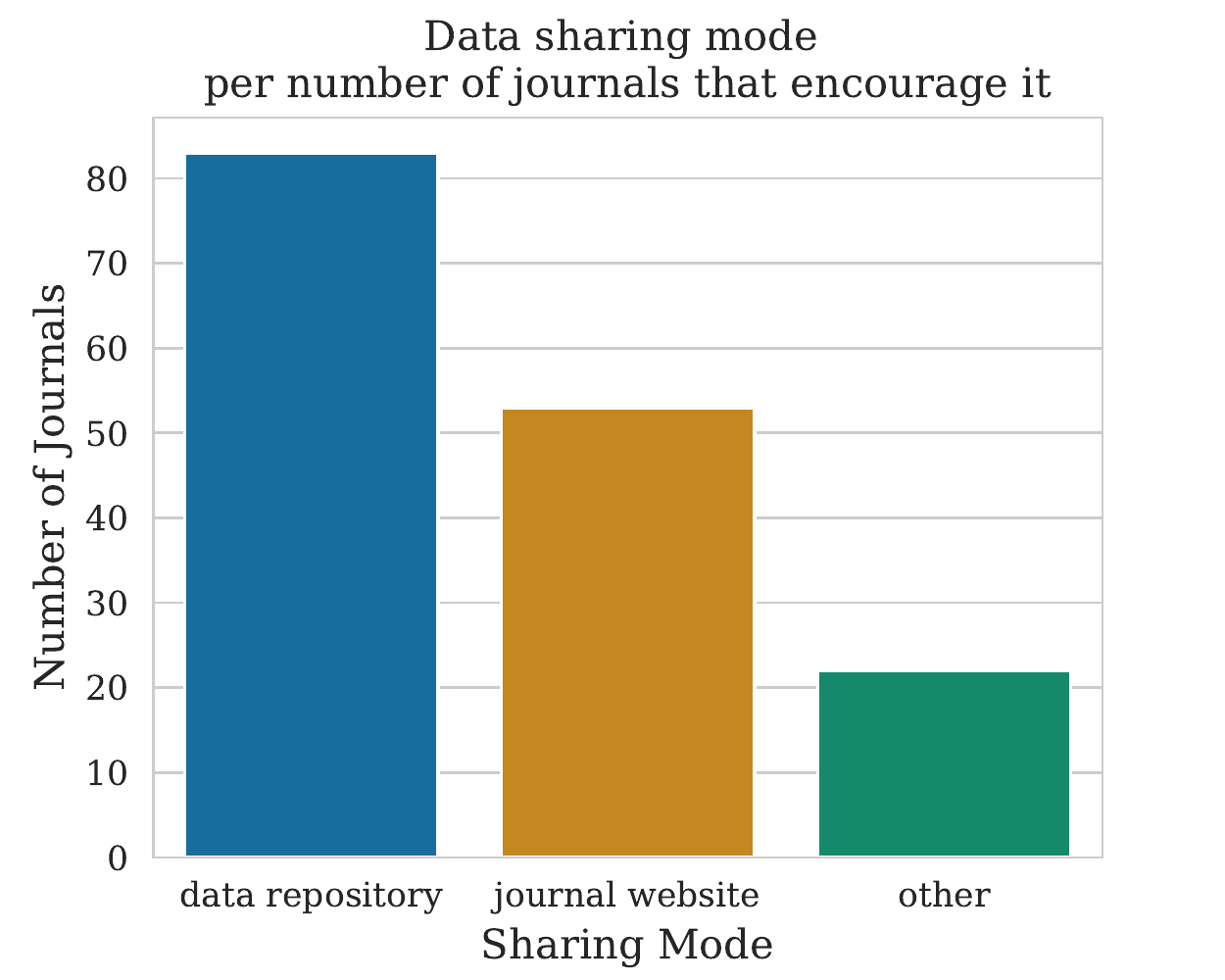}
  \caption{Aggregated results for most popular data sharing mode in anthropology, economics, history, PoliSci+IR, psychology, and sociology from Ref.~\cite{crosas2018data}.} \label{fig:mode}
\end{figure}

Some academic journals endorse this approach for publishing research outputs, and they often encourage (or require) their authors to release a replication package upon publication. Data repositories, such as Dataverse or Dryad, are the predominantly encouraged mode for sharing research data and code followed by journals' own websites (Figure~\ref{fig:mode})~\cite{crosas2018data}. For example, the American Journal of Political Science (AJPS) and the journal Political Analysis have their own collections within the Harvard Dataverse repository, which is their required venue for sharing research data and code. 

Recent case studies~\cite{collberg2016repeatability} reported that the research material published in data repositories does not often guarantee reproducibility. This is in part because, in the current form, data repositories do not capture all software and system dependencies necessary for code execution. Even when this information is documented by the original authors in an instructions file (like readme), contextual information still might be missing, which could make the process of research verification and reuse hard or impossible. This is also often the case with some of the alternative ways of publishing research data and code, for example, through the journal's website. A study~\cite{rowhani2018badges} reported that the majority of supplemental data deposited on a journal's website was inaccessible due to broken links. Such problems are less likely to happen in data repositories that follow standards for long-term archival and support persistent identifiers.

Some researchers prefer to release their data and code on their personal websites or websites like GitHub and GitLab. This approach does not natively provide a standardized persistent citation for referencing and accessing the research materials, nor sufficient metadata to make it discoverable in data search engines like Google Dataset Search and DataCite search. In addition, it does not guarantee long-term accessibility as data repositories do. Because research deposited this way does not typically contain a runtime environment, nor system or contextual information, this approach is also often ineffective in enabling computational reproducibility~\cite{pimentel2019large}. 

New cloud services have emerged to support research data organization, collaborative work, and reproducibility~\cite{perkel2019make}. Even though the number of different and useful reproducibility tools is constantly increasing, in this paper, we are going to focus on the following projects: Code Ocean~\cite{staubitz2016codeocean}, Whole Tale~\cite{brinckman2019computing}, Renku~\footnote{https://renkulab.io} and Binder~\cite{jupyter2018binder, kluyver2016jupyter}. All of these tools are available through a web browser, and they are based on the containerization technology Docker, which provides a standardized way to capture the computational environment that can be shared, reproduced, and reused.
\begin{enumerate}
    \item Code Ocean is a research collaboration platform that enables its users to develop, execute, share, and publish their data and code. The platform supports a large number of programming languages including R, C/C++, Java, Python, and it is currently the only platform that supports code sharing in proprietary software like MATLAB and Stata.
    \item Whole Tale is a free and open-source reproducibility platform that, by capturing data, code, and a complete software environment, enables researchers to examine, transform and republish research data that was used in an academic article.
    \item Renku is a project similar to Whole Tale that focuses on employing tools for best coding practices to facilitate collaborative work and reproducibility. 
    \item  Binder is a free and open-source project that allows users to run notebooks (Jupyter or R) and other code files by creating a containerized environment using configuration files within a replication package (or a repository).
\end{enumerate}

Even though virtual containers are currently considered the most comprehensive way to preserve computational research~\cite{piccolo2016tools,jimenez2015role}, they do not entirely comply with modern scientific workflows and needs. Through the use of containers, the reproducibility platforms in most cases fail to support FAIR principles (Findable, Accessible, Interoperable, and Reusable)~\cite{wilkinson2016fair}, standardized persistent citation, and long-term preservation of research outputs as data repositories strive to do. Findability is enabled through standard or community-used metadata schemas that document research artifacts. Data and code stored in a Docker container on a reproducibility platform are not easily visible nor accessible from outside of the container, which thus hinders their findability. This could be an issue for a researcher looking for a specific dataset rather than a replication package. In addition, unlike data repositories, reproducibility platforms do not undertake a commitment to the archival of research materials. This means that, for example, in a scenario where a reproducibility platform runs out of funding, the deposited research could be inaccessible. 

Individually data repositories and reproducibility platforms cannot fully support scientific workflows and requirements for reproducibility and preservation. This paper explains how these shortcomings could be overcome through integration that would result in a robust paradigm for preserving computational research and enabling reproducibility and reuse while making the replication packages FAIR. We argue that through the integration of these existing projects, rather than inventing new ones, we could combine the functionalities that effectively complement each other.



\section{Related work}

Through integration, reproducibility platforms and data repositories create a synergy that addresses weaknesses of both approaches. Some of these integrations are already on the way:

\begin{itemize}
    \item CLOCKSS~\footnote{https://clockss.org} is an archiving repository that preserves data with regular validity checks. Unlike other data repositories, it does not provide public or user access to the preserved content, except in special cases that are referred to as ``triggered content''. Code Ocean has partnered with CLOCKSS to preserve in perpetuity research capsules associated with publications from some of the collaborating journals.
    \item The Whole Tale platform relies on integrations with external resources for long-term stewardship and preservation. They already enable data import from data repositories, and a publishing functionality for a replication package is currently underway through DataONE, Dataverse, and Zenodo~\cite{chard2019implementing}.
    \item Stencila is an open-source office suite designed for creating interactive, data-driven publications. With its familiar user interface, it is geared toward the users of Microsoft Word and Excel. It integrates data and code as a self-contained part of the publication, and it also enables external researches to explore the data and write custom code. Stencila and the journal eLife have partnered up to facilitate reproducible publications~\cite{maciocci2019introducing}.
\end{itemize}

\section{Implementation}

In this paper, we present our developments in the context of the Dataverse Project, which is a free and open-source software platform to archive, share, and cite research data. Currently, 55 institutions around the globe run Dataverse instances as their data repository. 

Dataverse's integration with the reproducibility platforms has propelled a series of questions and developments around advancing reproducibility for its vast and diverse user community. First, while container files can be uploaded to Dataverse, there is no special handling for these files, which can result in mixed outcomes for researchers trying to verify reproducibility. Second, it is important to facilitate the capture of computational dependencies for the Dataverse users who choose not to use a reproducibility platform. Finally, in a replication package with multiple seemingly disorganized code files, it would be important to minimize the time and effort of an external user who wants to rerun and reuse the files. Therefore, new functionality to support container-based deposits, organization, and access needs to be added to Dataverse to improve reproducibility. 

\subsection{Integration with reproducibility platforms}

Dataverse integration with the reproducibility platforms should allow both adding new research material into Dataverse, and importing and reusing the existing material from Dataverse into a reproducibility platform. This communication is implemented through a series of existing and new APIs. The reproducibility platforms that have an ongoing integration collaboration with Dataverse are Code Ocean, Whole Tale, Binder, and Renku.

\begin{figure}
  \centering
    \includegraphics[width=\linewidth]{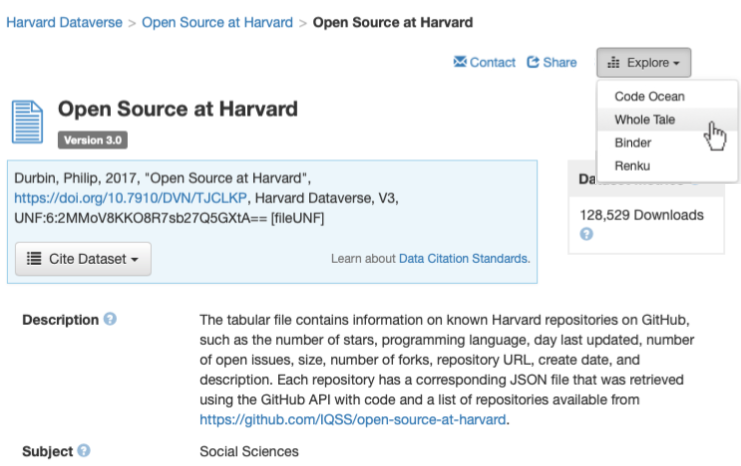}
   \caption{Preliminary view of how research stored in Dataverse can be viewed and explored in reproducibility platforms with a button click.} \label{fig:dataverse}
\end{figure}

Importing research material from Dataverse means that data and code that already exist in Dataverse could be transferred directly into a reproducibility platform. On the Dataverse side, this is implemented through a new button "Explore", shown in Figure~\ref{fig:dataverse}. When the button is clicked, the replication package is copied and sent to a reproducibility platform where it, using the configuration files from the package, creates a Docker container, places all data and code into it and provides a view through a web browser. This means that the Dataverse users will not need to download any of the files to their personal computers, nor will they need to set up a computational environment to execute and explore the deposited files. So far, the "Explore" button is functional for the Whole Tale platform, while the others are underway.~\footnote{Dataverse documentation for integrations: http://guides.dataverse.org/en/4.20/admin/ external-tools.html}

\begin{figure}
    \centering
    \includegraphics[width=\linewidth]{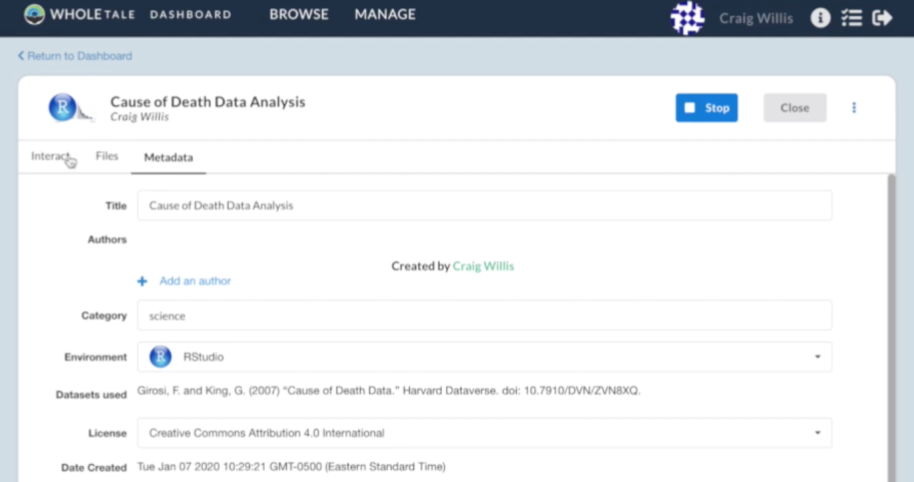}
  \caption{Using data from Dataverse in the Whole Tale environment. Snapshot from YouTube video \url{https://www.youtube.com/watch?v=oWEcFpEUmrU}. Credit: Craig Willis.} \label{fig:wt}
\end{figure}

\begin{figure}
    \centering
    \includegraphics[width=\linewidth]{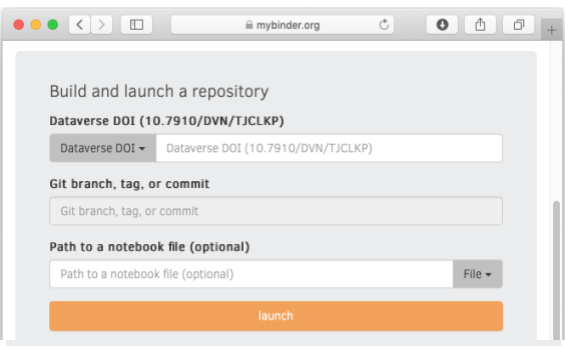}
  \caption{Binder’s GUI (\url{https://mybinder.org}) supports viewing content from Dataverse.} \label{fig:binder}
\end{figure}

The researcher whose starting point is the reproducibility platform will be able to import materials for their analysis directly from Dataverse. An example where a researcher is importing Dataverse open data as "external data" into Whole Tale is shown in Figure~\ref{fig:wt}, as this integration is now implemented. Similarly, Figure~\ref{fig:binder} shows new integration developments with the lightweight cloud platform, Binder, that now enables the users to import and view data from Dataverse. The export of the encapsulated research material into Dataverse will also be possible, which means that, once an analysis is ready for dissemination, the researchers would initiate "analysis export" in a reproducibility platform, that would then copy the files from a Docker container into Dataverse. This way, all necessary computational dependencies are automatically recorded by a reproducibility tool and stored at a data repository following preservation standards. This functionality is already implemented on Renku.~\footnote{Integration code at https://github.com/SwissDataScienceCenter/renku-python/pull/909}


\subsection{Handling containers}

Importing replication packages from the reproducibility platforms means that Dataverse would need to support the capture of their virtual containers. Since all aforementioned reproducibility platforms are based on the containerization technology Docker, new Dataverse developments focus on Docker containers. Docker containers can be built automatically from the instructions laid out in a "Dockerfile". A Dockerfile is an often tiny text file that contains commands, typically for installing software and dependencies, to set up a runtime environment needed for research analysis. The Dataverse platform will encourage depositing Dockerfiles to capture the computational environment. This would allow the users to explore replication packages in any supported reproducibility tool as Dockerfiles are agnostic to computational platforms. An alternative solution would be to create a Dataverse Docker registry where the whole images would be preserved. This approach will not be pursued at the time due to excessive storage requirements.

It is important to mention that at present, any file can be stored at Dataverse, including a Dockerfile, and that there are currently dozens of Dockerfiles stored at Harvard's instance of Dataverse. However, whereas previously Dockerfiles were considered as "other files", in the new development they will be pre-identified at upload, and thus will require additional metadata. When a reproducibility platform automatically generates a Dockerfile, it is likely to be suitable for portability and preservation. However, when a researcher prepares it, this might not be the case. Dockerfiles could be susceptible to some of the practices that cause irreproducibility, like the use of absolute (fixed) file paths, which is why Dataverse will encourage its users to use best practices when depositing these files. In particular, before depositing a Dockerfile, the researcher will be prompted to confirm that their Dockerfile does not include any of the common reproducibility errors.

\subsection{Capturing execution commands}

In addition to capturing a Docker container via Dockerfile, it is important to capture the sequence of steps that the user ran to obtain their results. This applies to the results obtained with command-line languages such as Python, MATLAB, Julia. Capturing the command sequence is particularly important when there are multiple code files within the replication dataset without the clear notation in which order they should be executed. The commands will be captured in the replication package metadata using a community standard to be determined (see, for example, RO-Crate~\cite{carragainro}).

In case the replication package was pulled from a reproducibility tool, like Code Ocean and Whole Tale, these replication commands would be automatically populated. For example, Code Ocean generates the commands that build and run a Docker container for each replication package, and it also encourages researchers to specify the command sequence in a file called "run" to automatize their code. This means that all command sequences that run "outside" and "inside" the container are captured. Dataverse users who choose not to use a reproducibility platform would need to manually specify this sequence based on presented best practices.

\subsection{Improving FAIR-ness}

Because no dataset would be 'hidden' within a virtual container at Dataverse, all files originally used in research would be indexed and thus findable by one of the common dataset search engines. They would also be accessible directly from the dataset landing page on the web. Their interoperability and reusability would be now improved with the integration with the reproducibility platforms, as the barriers to creating a runtime environment and running code files would be alleviated. 

\begin{figure*}[htb!]
\centering
   \includegraphics[width=0.85\linewidth]{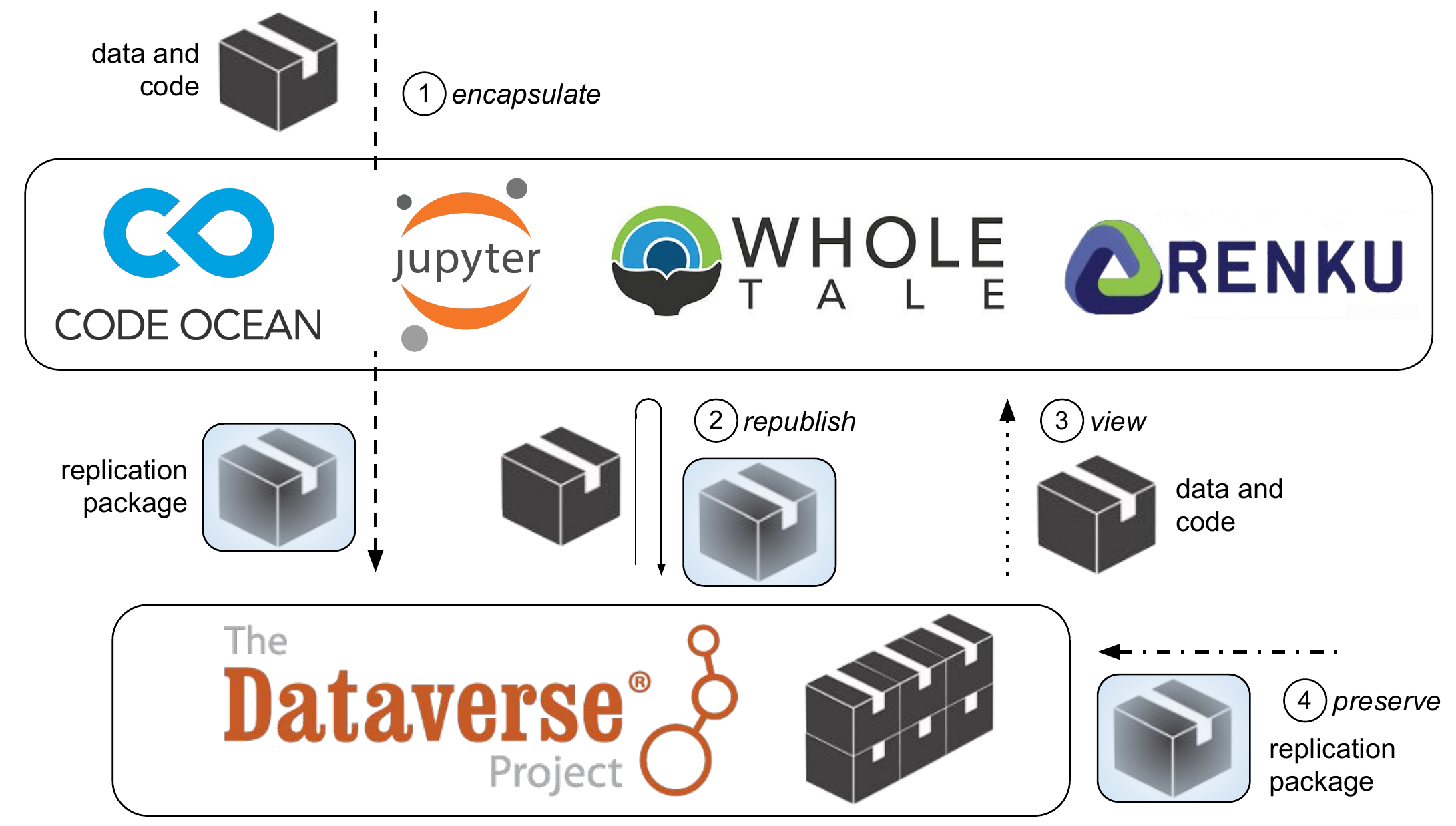}
   \caption{Four main workflows that Dataverse aims to support with reproducibility platform integration.} \label{fig:int}
\end{figure*}

\subsection{New data metrics}

Dataverse traditionally aims to provide incentives to researchers to share data through data citation credit, data metrics such as a count of downloads for datasets and access requests for restricted data. One of the completed new developments includes integrating certifications or science badges, such as Open Data and Open Materials, within a dataset landing page on Dataverse.

The new support for reproducibility tools and containers will also result in creating new metrics for the users. The datasets that are deposited through a reproducibility platform into Dataverse will be denoted with a 'reproducibility certification' badge that will signal their origin and easy execution on the cloud. For example, a replication package that was received from a reproducibility platform Whole Tale, will include its origin information and encourage its exploration and reuse through Whole Tale. 

\section{Functionality and use-cases}

The Dataverse integration with the reproducibility platforms and the new developments that improve the reproducibility of deposited research will facilitate research workflows relating to verification, preservation, and reuse in the following ways (shown in Figure~\ref{fig:int}):

\begin{enumerate}
    \item Research encapsulation. The first supported workflow enables authors to deposit their data and code through Code Ocean, Whole Tale, or Renku, which then create a replication package that is sent for dissemination and preservation to Dataverse. The Dataverse users who were not previously familiar with the containerization technology Docker will now be able to containerize their research through the new workflow. In addition, this workflow is particularly important for prestigious academic journals that verify research reproducibility through third-party curation services and a reproducibility platform. For example, code review at the journal Political Analysis, which collaborates with Code Ocean and Harvard Dataverse for data dissemination and preservation, will be significantly sped up with the deployment of this workflow, as all the code associated with a publication will already be automatized, containerized and available on the cloud.

    \item Modify and republish research. The second workflow covers pulling a replication package from Dataverse and republishing it after an update. This would create a new version of the package in Dataverse, as well as track provenance about the original package. Peer-review and revisions of the package should thus be much easier. In addition, the replication packages on Dataverse that currently do not have information on their runtime environment could be updated and republished with a Dockerfile generated by one of the reproducibility platforms.

    \item View deposited research materials. The third functionality allows viewing and exploring the content of deposited research without the need to download the files and install new software. This could be particularly valuable for external researchers and students who would like to understand research results or reuse data or code. 
    
    \item Preserve computational environment with Dockerfile. Through the new developments in Dataverse that encourage depositing Dockerfiles with best practices, the researchers who are experienced in using Docker will now be able to adequately preserve these files in the repository.

\end{enumerate}

\section{Conclusions}

In the last decade, there has been extensive discussion around preservation, reproducibility, and openness of computational research, which resulted in creating multiple new tools to facilitate these efforts, the most popular being data repositories and reproducibility platforms. However, individually these two approaches cannot fully facilitate findable, interoperable, reusable, and reproducible research materials. This paper presents a robust solution achieved through their integration. 

Described integrations have resulted in developing new functionality in Dataverse, such as expanding on the existing API, introducing new replication-package metadata, and handling virtual containers via Dockerfile. In addition to allowing research preservation in a reproducible and reusable way through the integrations, Dataverse aims to identify new and useful data metrics to be displayed on the dataset landing page. Due to the fact that there is an increasing number of similar reproducibility tools, this paper also advocates for considering integration with an existing solution before (re)inventing a new reproducibility tool.